\documentclass[a4paper,12pt]{article}
\usepackage{graphicx}
\usepackage{amssymb,amsthm}
\usepackage{color}
\title{Bistable firing pattern in \\a neural network model}

\begin{document}
		\maketitle
\noindent P. R. Protachevicz$^1$, F. S. Borges$^2$, E. L. Lameu$^3$, P.
Ji$^{4,5}$, K. C. Iarosz$^6$, A. H. Kihara$^2$, I. L. Caldas$^6$, J. D. Szezech
Jr.$^{1,7}$, M. S. Baptista$^8$, E. E. N. Macau$^3$, C. G. Antonopoulos$^9$,
A. M. Batista$^{1,7}$, J. Kurths$^{10,11,*}$\\
 
{\small
\noindent$^1$Graduate in Science Program - Physics, State University of
Ponta Grossa, PR, Brazil\\
$^2$Center for Mathematics, Computation, and Cognition, Federal University of
ABC, S\~ao Bernardo do Campo, SP, Brazil\\
$^3$National Institute for Space Research, S\~ao Jos\'e dos Campos, SP, Brazil\\
$^4$Institute of Science and Technology for Brain-Inspired Intelligence, Fudan
University,  Shanghai, China\\
$^5$Key Laboratory of Computational Neuroscience and Brain-Inspired
Intelligence (Fudan University), Ministry of Education, China\\
$^6$Institute of Physics, University of S\~ao Paulo, S\~ao Paulo, SP, Brazil\\
$^7$Department of Mathematics and Statistics, State University of Ponta Grossa,
Ponta Grossa, PR, Brazil\\
$^8$Institute for Complex Systems and Mathematical Biology, SUPA, University of
Aberdeen, Aberdeen, AB24 3UE, Scotland, United Kingdom\\
$^9$Department of Mathematical Sciences, University of Essex, Wivenhoe Park,
UK\\
$^{10}$Potsdam Institute for Climate Impact Research, Potsdam, Germany\\
$^{11}$Department of Physics, Humboldt University, Berlin, Germany
J\"urgen Kurths
kurths@pik-potsdam.de
}


\section*{Abstract}
Excessively high, neural synchronisation has been associated with epileptic
seizures, one of the most common brain diseases worldwide. A better
understanding of neural synchronisation mechanisms can thus help control or
even treat epilepsy. In this paper, we study neural synchronisation in a random
network where nodes are neurons with excitatory and inhibitory synapses, and
neural activity for each node is provided by the adaptive exponential
integrate-and-fire model. In this framework, we verify that the decrease in the
influence of inhibition can generate synchronisation originating from a pattern
of desynchronised spikes. The transition from desynchronous spikes to
synchronous bursts of activity, induced by varying the synaptic coupling,
emerges in a hysteresis loop due to bistability where abnormal (excessively
high synchronous) regimes exist. We verify that, for parameters in the
bistability regime, a square current pulse can trigger excessively high
(abnormal) synchronisation, a process that can reproduce features of epileptic
seizures. Then, we show that it is possible to suppress such abnormal
synchronisation by applying a small-amplitude external current on less than
10\% of the neurons in the network. Our results demonstrate that external
electrical stimulation not only can trigger synchronous behaviour, but more
importantly, it can be used as a means to reduce abnormal synchronisation and
thus, control or treat effectively epileptic seizures.\\

\noindent Keywords:  bistable regime, network, adaptive
exponential integrate-and-fire neural model, neural dynamics, synchronisation,
epilepsy


\section{Introduction}

Epilepsy is a brain disease that causes seizures and sometimes loss of
consciousness \cite{Chen2014,Chen2015}. Epileptic seizures are associated with
excessively high synchronous activities \cite{Li2007,Jiruska2013,Wu2015} of
neocortex regions or other neural populations
\cite{Fisher2005,Sierra-Paredes2007,Engel2013,Geier2017,Fisher2018}.
Electroencephalography has been used to identify and classify seizures
\cite{Noachtar2009}, as well as to understand epileptic seizures
\cite{Scharfman2014}. Abnormal activities have a short period of time, lasting
from a few seconds to minutes \cite{Trinka2015}, and they can occur in small or
larger regions in the brain \cite{McCandless2012,Kramer2012}. Two suggested
mechanisms responsible for the generation of partial epilepsy are the decrease
of inhibition and increase of excitation \cite{McCandless2012}. In experiments
and simulations, the reduction of excitatory and the increase of inhibitory
influence have been effective in suppressing and preventing synchronised
behaviours \cite{Traub1993,Schindler2008}. Traub and Wong \cite{Traub1982}
showed that synchronised bursts that appear in epileptic seizures depend on
neural dynamics.

Single seizures can not kill neurons, however recurrent ones can do so and
thus, can lead to chronic epilepsy \cite{Dingledine2014}. Evidence that
supports this further is provided by abnormal anatomical alterations, such as
mossy fiber sprouting \cite{Danzer2017}, dendritic reconfigurations
\cite{Wong2005,Wong2008}, and neurogenesis \cite{Jessberger2015,Cho2015}.
In fact, such alterations change the balance between inhibition and excitation
\cite{Holt2013,Silva2003}. Wang et al. \cite{Wang2017} demonstrated that a
small alteration in the network topology can induce a bistable state with an
abrupt transition to synchronisation. Some {\it in vitro} seizures generated
epileptiform activities when inhibitory synapses were blocked or excitatory
synapses were enhanced \cite{Traub1994,White2002}. Several studies showed that
epileptiform activities are related not only with unbalanced neural networks,
but also with highly synchronous regimes \cite{Uhlhaas2006,Abdullahi2017}.

Different routes to epileptic seizures were reported by Silva et al.
\cite{Silva2003}. The authors considered epilepsy as a dynamical disease and
presented a theoretical framework where epileptic seizures occur in neural
networks that exhibit bistable dynamics. In the bistable state, transitions can
happen between desynchronous and synchronous behaviours. Suffczynski et al.
\cite{Suffczynski2004} modelled the dynamics of epileptic phenomena by means
of a bistable network. 

Many works reported that periodic electrical pulse stimulation facilitates
synchronisation, while random stimulation promotes desynchronisation in networks
\cite{Cota2009}. Electrical stimulation can be applied in different brain
areas, for instance in the hippocampus, thalamus, and cerebellum
\cite{McCandless2012}. The mechanism for electrical stimulation to cease
seizures is still not completely understood, however, signal parameters such as
frequency, duration and amplitude can be changed to improve the efficiency of
the treatment of epilepsy \cite{McCandless2012}. The electrical stimulation
has been used as an efficient treatment for epilepsy in the hippocampus
\cite{Velasco2007}. In \cite{Antonopoulos2016}, the author studied external
electrical perturbations and their responses in the brain dynamic network of the
Caenorhabditis elegans soil worm. It was shown that when one perturbs specific
communities, keeping the others unperturbed, the external stimulations
propagate to some but not all of them. It was also found that there are
perturbations that do not trigger any response at all and that this depends on
the initially perturbed community.

Neural network models have been used to mimic phenomena related to neural
activities in the brain. Guo et al. \cite{Guo2016a} built a network model
where the postsynaptic neuron receives input from excitatory presynaptic
neurons. They incorporated autaptic coupling \cite{Guo2016b} in a biophysical
model. Delayed models have been considered in biological systems
\cite{Khajanchi2018}, for instance, Sun et al. \cite{Sun2018} analysed the
influence of time delay in neuronal networks. They showed that intra- and
inter-time delays can induce fast regular firings in clustered networks. In
this work, we build a random network with neural dynamics to study
synchronisation induced in a bistable state which is related to epileptic
seizures. In particular, we consider a network composed of adaptive exponential
integrate-and-fire (AEIF) neurons coupled by means of inhibitory and excitatory
synapses. The AEIF model mimics phenomenological behaviours of neurons
\cite{Clopath2006} and is appropriate to study even large networks
\cite{Naud2008}. Borges et al. \cite{Borges2017} verified that depending on
the excitatory synaptic strength and connection probability, a random network
of coupled AEIF neurons can exhibit transitions between desynchronised spikes
and synchronised bursts \cite{Protachevicz2018}. In the network considered
here, we observe the existence of bistability when it is unbalanced, namely
that the decrease of synaptic inhibition induces a bistable state. We analyse
the effects of the application of external square current pulses (SCP) by
perturbing the neural dynamics on the network using parameters that lead to a
bistable state, such as the excitatory and inhibitory synaptic conductances. We
find that, depending on the duration and amplitude of the external current, SCP
can either trigger or suppress synchronisation in the bistability region, an
idea that can be used further to treat epilepsy by suppressing excessive
synchronisation in affected brain regions.


\section{Methods}

\subsection{Neural network model}

We build a random network of $N=1000$ adaptive exponential
integrate-and-fire neurons \cite{Brette2005} with probability $p$ for the
formation of connections among them equal to 0.1 . The network consists of 80\%
excitatory and 20\% inhibitory neurons \cite{noback05}. The dynamics of each
neuron $i,\;i=1,\ldots,N$ in the network is given by the set of equations
\begin{eqnarray}
C_{\rm m}\frac{d V_i}{d t} & = & - g_L(V_i - E_L) + g_L {\Delta}_T \exp 
\left(\frac{V_i - V_T}{{\Delta}_T} \right)\nonumber \\ \nonumber
& + & I_i - w_i +    \sum_{j=1}^{N} (V_{\rm{REV}}^j - V_i) M_{ij}g_{j} 
+ \Gamma_i, \nonumber \\ 
\tau_w \frac{d w_i}{d t} & = & a_i (V_i - E_L) - w_i,\\ \nonumber 
\tau_s \frac{dg_{i}}{dt} & = & -g_{i}.
\label{eqIFrede}
\end{eqnarray}
The membrane potential $V_i$ and adaptation current $w_i$ represent the state
of each neuron $i$. The capacitance membrane $C_{\rm m}$ is set to
$C_{\rm m}=200$pF, the leak conductance to $g_{L}=12$ns, the resting potential to
$E_L=-70$mV, the slope factor to $\Delta_T=2.0$mV and the spike threshold to
$V_T=-50$mV. The adaptation current depends on the adaptation time constant
$\tau_w=300$ms and the level of subthreshold adaptation $a_i$ that is randomly
distributed in the interval $[0.19,0.21]$ns. We consider the injection of
current $I_i$ to each neuron $i$ in terms of the relative rheobase current
$r_i=I_i/I_{\rm rheobase}$ \cite{Naud2008}. The rheobase is the minimum amplitude
of the applied current to generate a single or successive firings. The
application of this constant current allows neurons to change their potentials
from resting potentials to spikes. The value of the rheobase depends on the
neuron parameters. The external current arriving at neuron $i$ is
represented by $\Gamma_i$. We consider the external current according to a SCP
with amplitude $A_I$ and time duration $T_I$. The random connections in the
network are described by the binary adjacency matrix $M_{ij}$ with entries
either equal to 1 when there is a connection from $i$ to $j$ or 0 in the
absence of such a connection. $g_i$ is the synaptic conductance, $\tau_s$
the synaptic time constant, and $V_{\rm{REV}}$ the synaptic reversal
potential. We consider $\tau_s=2.728$ms, $V_{\rm REV}=0$mV for excitatory
synapses, and $V_{\rm REV}=-80$mV for inhibitory synapses. The synaptic
conductance decays exponential with a synaptic time constant $\tau_s$. When the
membrane potential of neuron $i$ is above the threshold $V_i>V_{\rm thres}$
\cite{Naud2008}, the state variable is updated by the rule
\begin{eqnarray}
V_i &\to & V_r =-58{\rm mV}, \nonumber\\
w_i &\to& w_i + 70{\rm pA},\\
g_{i} &\to& g_{i}+g_s, \nonumber
\end{eqnarray}
where $g_s$ assumes the value of $g_{\rm exc}$ when neuron $i$ is excitatory
($i\leq 0.8N$) and $g_{\rm inh}$ when neuron $i$ is inhibitory ($i>0.8N$). In
this work, we study the parameter space $(g_{\rm exc}, g_{\rm inh})$ and consider
a relative inhibitory synaptic conductance $g=g_{\rm inh}/g_{\rm exc}$. We consider
parameter values in which the individual uncoupled neurons perform spike
activities. The initial values of $V$ and $w$ are randomly distributed in the
interval $[-70,-50]$mV and $[0,70]$pA, respectively. The initial $g_i$ value is
equal to $0$.

\subsection{Synchronisation}

The synchronous behaviour in the network can be identified by means of
the complex phase order parameter \cite{Kuramoto1984} 
\begin{equation}
R(t)\exp({\rm i}\Phi(t))\equiv\frac{1}{N}\sum_{j=1}^{N}\exp({\rm i}\psi_{j}(t)),
\end{equation}
where $R(t)$ and $\Phi(t)$ are the amplitude and angle of a centroid phase
vector over time, respectively. The phase of neuron $j$ is obtained by
means of  
\begin{equation}
\psi_{j}(t)=2\pi m+2\pi\frac{t-t_{j,m}}{t_{j,m+1}-t_{j,m}},
\end{equation}
where $t_{j,m}$ corresponds to the time of the $m-$th spike of neuron $j$
($t_{j,m}< t < t_{j,m+1}$) \cite{Rosenblum96,Rosenblum97}. We consider that the
spike occurs for $V_j>V_{\rm thres}$. $R(t)$ is equal to $0$ for fully
desynchronised and $1$ for fully synchronised patterns, respectively.

We calculate the time-average order parameter $\overline{R}$
\cite{Batista2017} given by 
\begin{equation}
\overline{R}=\frac{1}{t_{\rm fin}- t_{\rm ini}} \int_{t_{\rm ini}}^{t_{\rm fin}} R(t)dt,
\end{equation} 
where $t_{\rm fin}-t_{\rm ini}$ is the time window. We consider $t_{\rm fin}$=200s 
and $t_{\rm ini}$=180s.

\subsection{Synaptic input}	

We monitor the instantaneous synaptic conductances arriving at each neuron $i$
through
\begin{equation}
I^{\rm ISC}_i(t) = \sum_{j=1}^N (V_{\rm REV}^j- V_i) M_{ij}g_{j} .
\end{equation}
The instantaneous synaptic input changes over time due to the excitatory and
inhibitory inputs received by neuron $i$. The average instantaneous synaptic
conductances is given by
\begin{equation}
I_{\rm syn}(t)=\frac{1}{N}\sum_{i=1}^{N} I^{\rm ISC}_i(t).
\end{equation}

\subsection{Coefficient of variation}

The $m-$th inter-spike interval ${\rm ISI}_i^m$ is defined as the difference
between two consecutive spikes of neuron $i$,
\begin{equation}
{\rm ISI}_i^m=t^{m+1}_i-t^m_i>0,
\end{equation}
where $t_i^m$ is the time of the $m-$th spike of neuron $i$. 

Using the mean value of ${\rm ISI}_i$, $\overline{\rm ISI}_i$, and its
standard deviation, $\sigma_{{\rm ISI}_i}$, we calculate the coefficient of
variation (CV) 
\begin{equation}
{\rm CV}_i = \frac{\sigma_{{\rm ISI}_i}}{\overline{\rm ISI}_i}.
\end{equation}
The average of ${\rm CV}$ ($\overline{\rm CV}$) is then obtained through
\begin{equation}
\overline{\rm CV}=\frac {1}{N}\sum_{i=1}^{N} {\rm CV}_i.
\end{equation}
Finally, we use $\overline{\rm CV}$ to identify spike (when
$\overline{\rm CV}<0.5$) and burst fire patterns (when
$\overline{\rm CV}\ge0.5$) \cite{Borges2017,Protachevicz2018}.

\subsection{Instantaneous and mean firing-rate}

The instantaneous firing-rate in intervals of $t_{\rm step}=1$ms is given by 
\begin{equation}
\label{eq5}
F(t)=\frac{1}{N}\sum_{i=1}^{N} \left(\int_{t}^{t + t_{\rm step}}\delta(t'-t_i)dt'
\right),
\end{equation}
where $t_i$ is the firing time of neuron $i$ in the time interval
$(t\le t_i\le t+1)$ms. This measure allows to obtain the instantaneous
population activity in the network. The mean firing-rate can then be calculated
by means of
\begin{equation}
\overline{F}=\frac{1}{\overline{\rm ISI}},
\end{equation}
where $\overline{\rm ISI}$ is the average ISI obtained over all $N$ neurons in
the network, that is
${\rm \overline{ISI}}=\frac{1}{N}\sum_{i=1}^N {\rm \overline{ISI}}_i$.


\section{Results}

\subsection{Inhibitory effect on synchronous behaviour}

\begin{figure}[hbt]
\centering
\includegraphics[scale=0.33]{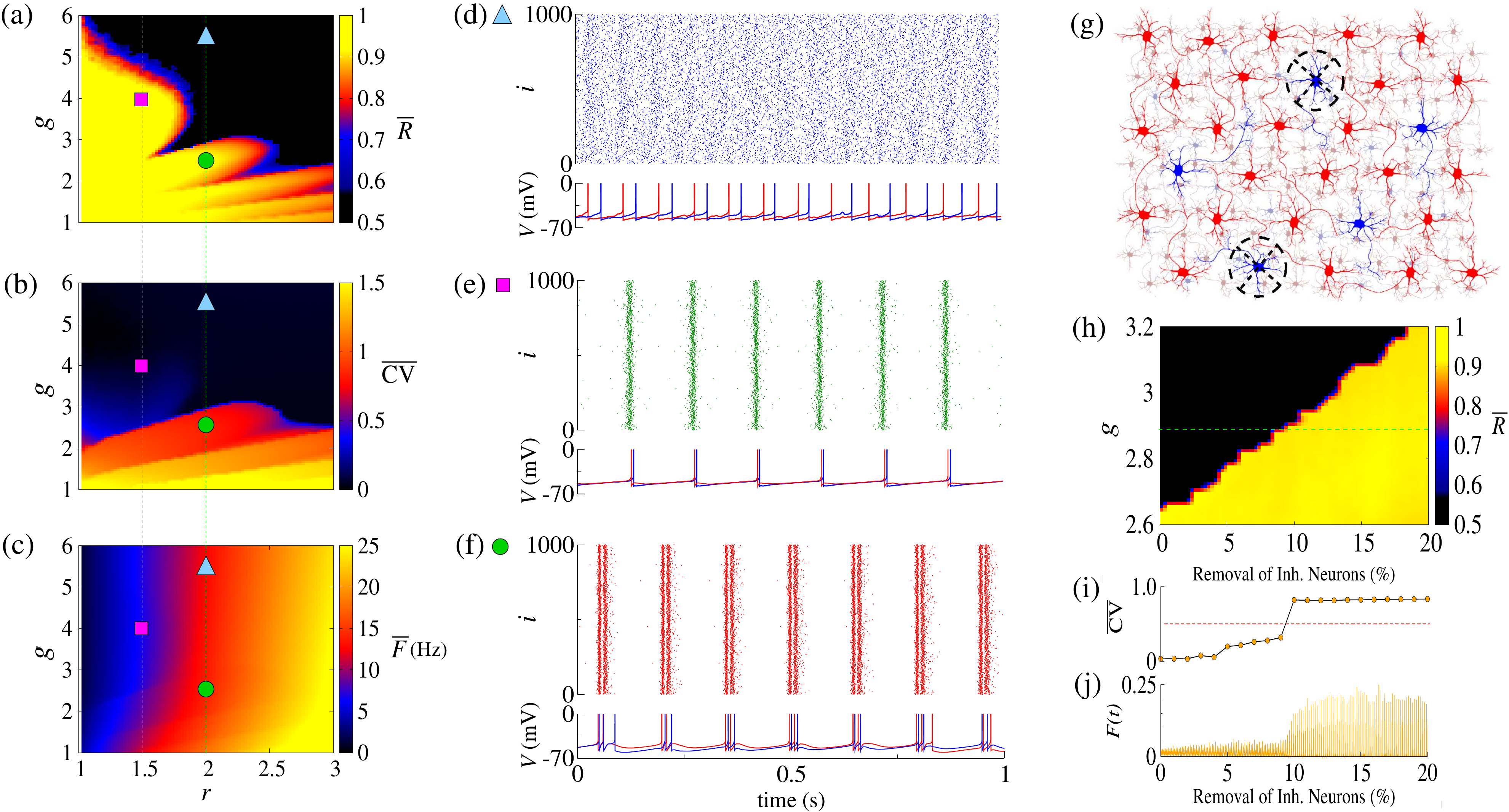}
\caption{(Colour online) Parameter space $(g,r)$ for the (a) time-average order
parameter ($\overline{R}$), (b) the mean coefficient of variation
($\overline{\rm CV}$), and (c) the mean firing-rate ($\overline{F}$). Raster
plot that displays the spiking activity over time and membrane potential are
shown for (d) desynchronised spikes for $r=2.0$ and $g=5.5$ (cyan triangle), (e)
synchronised spikes for $r=1.5$ and $g=4$ (magenta square), and (f)
synchronised bursts for $r=2.0$ and $g=2.5$ (green circle). Here, we consider
$g_{\rm exc}=0.4$ns. In panel (g), we illustrate a network composed of excitatory
(red) and inhibitory (blue) neurons, where some inhibitory neurons are removed
(black dashed circle). Figure (h) shows the time-average order parameter for
$g$ versus the percentage of inhibitory neurons removed from the network. The
green dashed line corresponds to $g=2.9$. The values of $\overline{\rm CV}$ and
instantaneous firing-rate are shown in panels (i) and (j), respectively.}
\label{fig1}
\end{figure}

The balance between excitation and inhibition generates an asynchronous
activity in the network \cite{Lundqvist2010,Ostojic2014}. However, for the
unbalanced network we observe synchronised spikes and bursts. Figures
\ref{fig1}(a), \ref{fig1}(b), and \ref{fig1}(c) show the time-average order
parameter (${\overline R}$), the mean coefficient of variation
($\overline{\rm CV}$) and the mean firing-rate (${\overline F}$), respectively,
for the parameter space $(g,r)$, where $g$ is the ratio between inhibitory
($g_{\rm inh}$) and excitatory ($g_{\rm exc}$) synaptic conductances, and $r$ the
relative rheobase current. For $g_{\rm exc}=0.4$ns and $g>6$, we observe that
$\overline{R}<0.5$ and that $\overline{\rm CV}<0.5$, corresponding to
desynchronised spikes. In Fig. \ref{fig1}(d), we see the raster plot and
membrane potential for 2 neurons in the network with a desynchronised
spike-pattern for $g=5.5$ and $r=2$ (blue triangle). For $g=4$ and $r=1.5$
(magenta square), the dynamics exhibits synchronised spikes (Fig.
\ref{fig1}(e)), as a result of setting $\overline{R}>0.9$ and
$\overline{\rm CV}<0.5$. Figure \ref{fig1}(f) shows synchronised bursts of
activity for $g=2.5$ and $r=2$ (green circle), where $\overline{R}>0.9$ and
$\overline{\rm CV}\geq 0.5$. Within this framework, we have verified the
existence of transitions from desynchronised spikes to synchronised bursting
activities without significant changes in the mean firing-rate.

The appearance of synchronous behaviour cannot only be related to the decrease 
of the inhibitory synaptic strength, but also to a loss of inhibitory neurons.
In particular, we show this in Fig. \ref{fig1}(g) which illustrates a network
composed of excitatory (red) and inhibitory (blue) neurons, where some
inhibitory neurons were removed (dashed circles). In Fig. \ref{fig1}(h), we see
that the synchronous behaviour depends on $g$ and the percentage of removed
inhibitory neurons. Figure \ref{fig1}(i) shows the transition from spiking
dynamics ($\overline{\rm CV}<0.5$) to bursting dynamics
($\overline{\rm CV}\geq 0.5$), and Fig. \ref{fig1}(j) shows the instantaneous
firing-rate $F(t)$. For $g=2.9$ and $g_{\rm exc}=0.4$ns (green dashed line), the
transition to synchronised bursts occurs when approximately 10\% of inhibitory
neurons are removed from the network, and as a consequence $F(t)$ reaches the
maximum value of $0.2$.

Concluding, alterations in the inhibitory synaptic strength or in the number of
inhibitory neurons can induce transition to synchronous patterns. Wang et al.
\cite{Wang2017} presented results where synchronisation transition occurs as a
result of small changes in the topology of the network, whereas here, we study
transitions caused due to changes in the inhibitory synaptic strength and
the emergence of a bistable regime.

\subsection{Bistable regime}

\begin{figure}[hbt]
\centering
\includegraphics[scale=0.22]{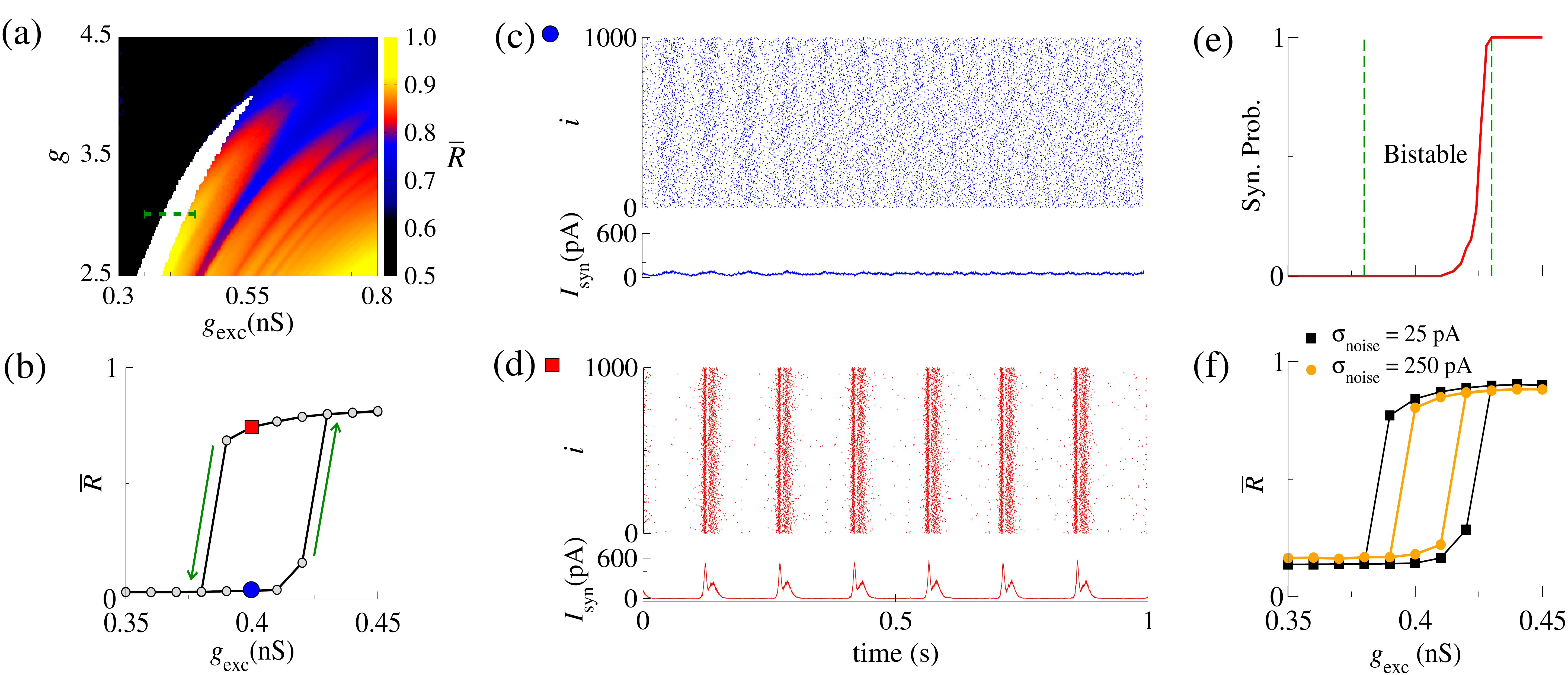}
\caption{(Colour online) (a) The parameter space $(g,g_{\rm exc})$ for $r=2$,
where $\overline{R}$ is encoded in colour. The black region corresponds to
desynchronised activity, whereas coloured regions indicate $\overline{R}>0.6$
and the white region represents the bistable regime. (b) The bistable region
indicated in the parameter space of (a) by means of a green dashed line. Panels
(c) and (d) show the raster plots and $I_{\rm syn}$ for desynchronised spikes
(blue circle) and synchronised bursts (red square), respectively. We identify
bistability by checking when
$\overline{R}_{\rm backward}-\overline{R}_{\rm forward}>0.4$ and consider two trials
for each set of parameter values. (e) The synchronisation probability as a
function of $g_{\rm exc}$. (f) $\overline{R}\times g_{\rm exc}$ for
$\sigma_{\rm noise}$ equal to $25$pA and $250$pA.}
\label{fig2}
\end{figure}

Next, we analyse synchronisation in the parameter space $(g,g_{\rm exc})$. In
particular, Fig. \ref{fig2}(a) shows ${\overline R}$ with values depicted in
the colour bar. The black region corresponds to desynchronised spike activity,
while the remaining coloured regions are associated with burst activities. The
white region represents the bistable regime, where desynchronised spikes or
synchronised bursts are possible depending on the initial conditions. In the
bistable regime, decreasing $g_{\rm exc}$ (backward direction), $\overline{R}$ is
higher than increasing $g_{\rm exc}$ (forward direction), as shown in Fig.
\ref{fig2}(b) for $g=3$, $r=2$, and $g_{\rm exc}=[0.35,0.45]$ns (green dashed
line in Fig. \ref{fig2}(a)). We identify bistability (white region) in the
parameter space when the condition
$\overline{R}_{\rm backward}-\overline{R}_{\rm forward}>0.4$ is fullfilled. The
raster plot and instantaneous synaptic input for desynchronised spikes (blue
circle) and synchronised bursts (red square) are shown in Figs. \ref{fig2}(c)
and \ref{fig2}(d), respectively. When the dynamics on the random network is
characterised by desynchronised spikes, the instantaneous synaptic inputs
exhibit $I_{\rm syn}(t)\approx50$pA. For synchronised bursts,
$I_{\rm syn}(t)\approx 0$ when a large number of neurons in the network are
silent (i.e. not firing), and $I_{\rm syn}(t)>200$pA during synchronous firing
activities. In Fig. \ref{fig2}(e), we compute the probability of occurrence of
excessively high synchronicity within the bistable regime. We observe a small
synchronisation probability value in the bistable region. This result has a
biological importance due to the fact that the seizure state is a relatively
small probability event compared with the normal state. DaQing et al.
\cite{DaQing2017} showed that noise can regulate seizure dynamics in partial
epilepsy. Figure \ref{fig2}(f) displays $\overline{R}\times g_{\rm exc}$ for
Gaussian noise with mean $0$ and standard deviation $\sigma_{\rm noise}$ equal to
$25$pA and $250$pA. We verify that the bistable region decreases when the noise
level increases.

\begin{figure}[hbt]
\centering
\includegraphics[scale=0.4]{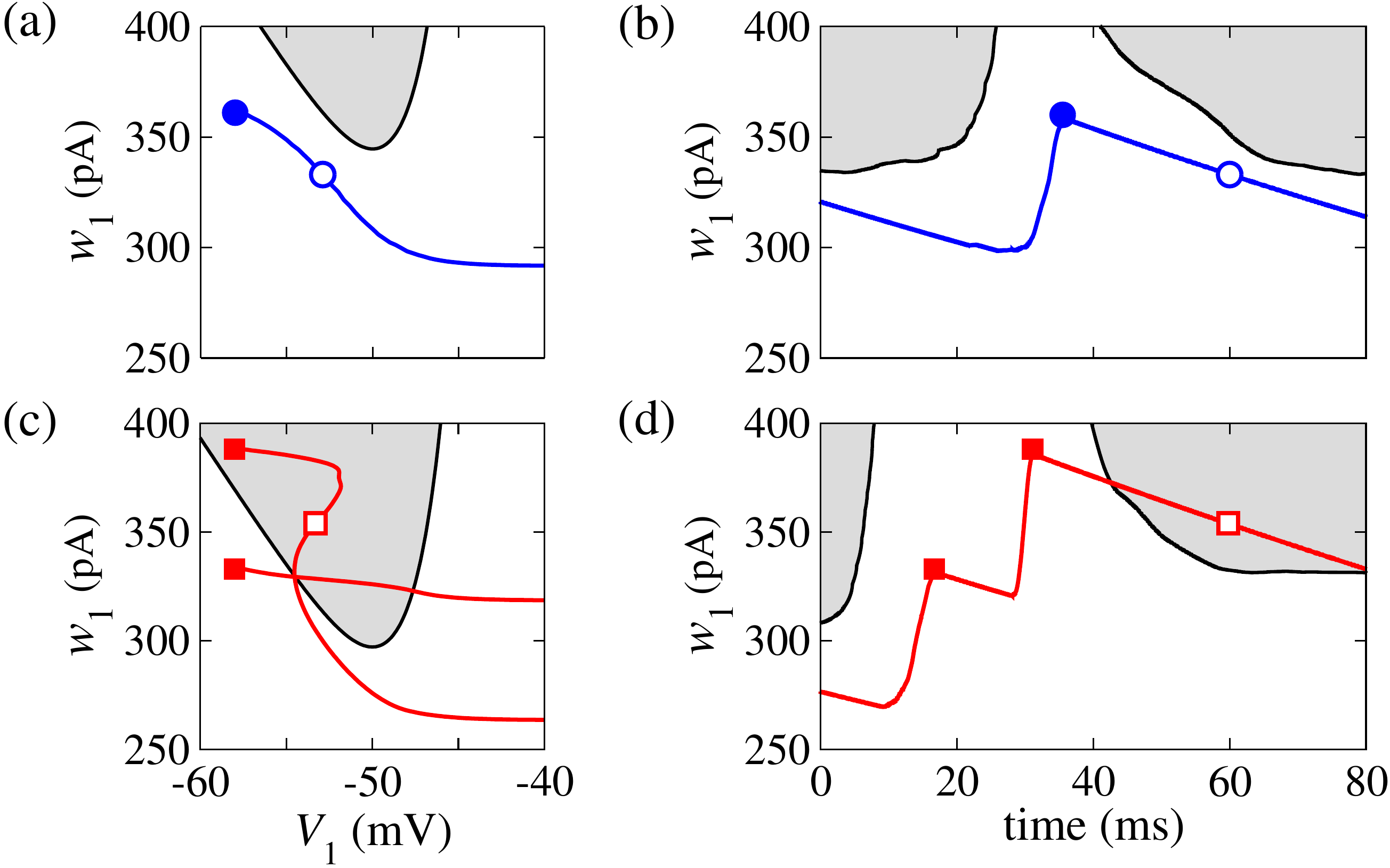}
\caption{(Colour online) Phase space $(w_1,V_1)$ ((a) and (c)) and time
evolution of $w_1$ ((b) and (d)) for spikes (blue) and burst activity (red).
The grey regions correspond to $dV_1/dt<0$ and the black line represents
$dV_1/dt=0$ ($V$-nullcline).}
\label{fig3}
\end{figure}

In the bistable regime, we investigate the evolution of a trajectory for a
finite time interval in the phase space $(w_i,V_i)$ and the time evolution
of $w_i$ shown in Fig. \ref{fig3} for $i=1$, where the grey regions
correspond to $dV_i/dt<0$. The boundary between the grey and white regions
(black line) is given by $dV_i/dt=0$, the $V_i$-nullcline \cite{Naud2008}.
During spiking activity, the trajectory (see Fig. \ref{fig3}(a)) and time
evolution of $w_i$ (see Fig. \ref{fig3}(b)) do not cross the $V_i$-nullcline.
For bursting activities (see Figs. \ref{fig3}(c) and \ref{fig3}(d)), we observe
that $w_i$ lies in the region enclosed by the $V_i$-nullcline. The emergence of
the bistable behaviour is related to changes in the $V_i$-nullcline caused by
the variation of $I_{\rm syn}$.

\subsection{External square current pulse}

Here, following a similar idea as in \cite{Antonopoulos2016}, we investigate
the effect of the application of SCP on the bistable regime. We apply SCP
considering different values of $A_I$, $T_I$, and number of removed inhibitory
neurons. The SCP is immediately switched off after $T_I$ and the analysis of
the effect on the dynamical behaviour is started.

\begin{figure}[hbt]
\centering
\includegraphics[scale=0.5]{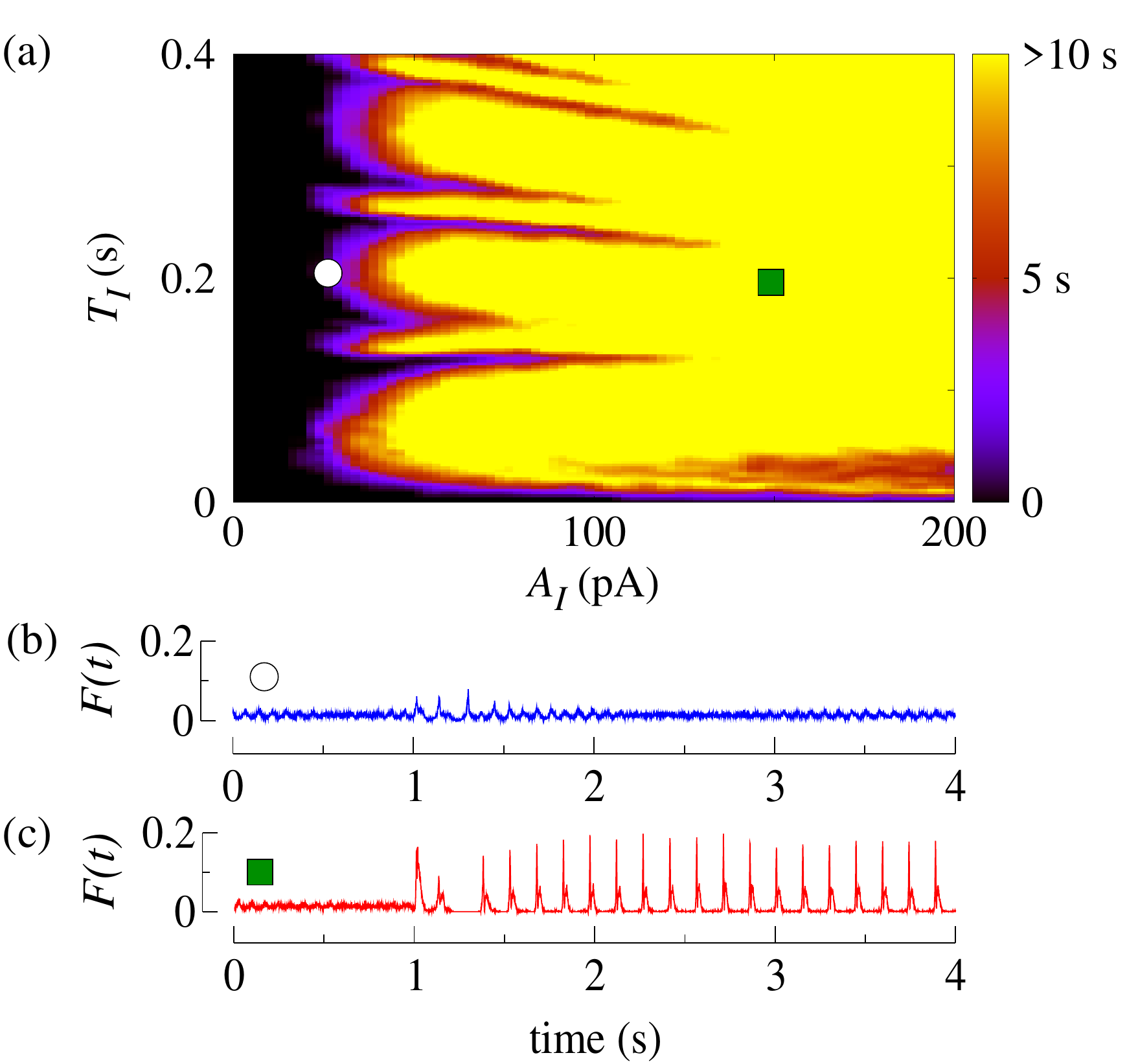}
\caption{(Colour online) (a) The parameter space $(T_I,A_I)$ in the bistable
regime, where the colour bar indicates the time the system shows synchronised
burst behaviour after the application of SCP. Instantaneous firing-rate for
values for (b) white circle ($A_I=25$pA, $T_I=0.2$s) and (c) green square
($A_I=150$pA, $T_I=0.2$s). Note that in this figure $g_{\rm exc}=0.4$ns, $g=3$
and $r=2$.}
\label{fig4}
\end{figure}

\begin{figure}[hbt]
\centering
\includegraphics[scale=0.41]{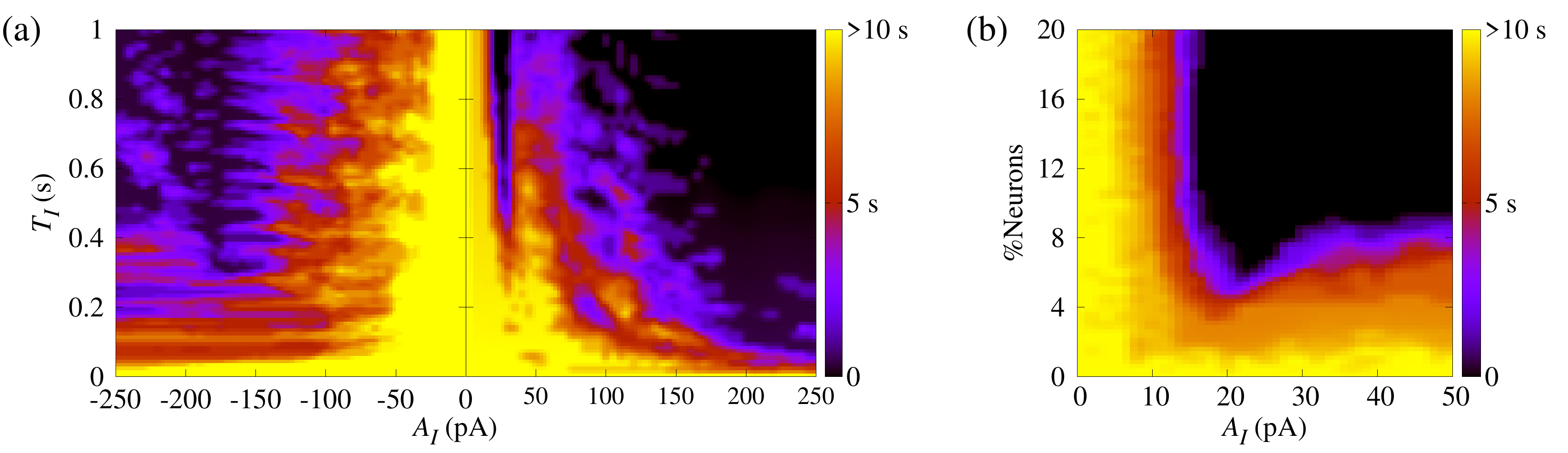}
\caption{(Colour online) (a) The parameter space $(T_I,A_I)$, where the colour
bar indicates the time the system shows synchronised burst behaviour after the
application of SCP. (b) Number of perturbed neurons as a function of $A_I$.
Note that in this figure we consider $g_{\rm exc}=0.4$ns, $g=3$ and $r=2$.}
\label{fig5}
\end{figure}

Initially, we apply SCP to all neurons in the network with parameter values
in the bistable regime with desynchronous behaviour (white region in Fig.
\ref{fig2}(a)). Figure \ref{fig4}(a) displays the time (in colour scale) that
the neurons show a synchronised pattern after the application of SCP. In the
black region, we see that SCP does not change the dynamical behaviour, namely
the neurons remain in a regime of desynchronised behaviour. The yellow region
depicts the values of $T_I$ and $A_I$ of the SCP that induce a change in the
behaviour of the neurons from desynchronised spikes to synchronised bursts.
Picking up one point close to the border of the black and blue regions (white
circle), we see that the instantaneous firing-rate ($F(t)$) of Eq. \ref{eq5}
(see Fig. \ref{fig4}(b), blue line) exhibits low-amplitude oscillations
corresponding to desynchronised spikes. For $T_I$ and $A_I$ values in the
yellow region, $F(t)$ (see Fig. \ref{fig4}(c), red line) exhibits a
high-amplitude oscillation after the application of SCP, corresponding to
synchronised bursts. For sufficiently large amplitudes, the change in the
behaviour induced by SCP does not depend on time. Importantly, perturbations
with small amplitudes applied for short times is a sufficient condition for the
induction of synchronous burst activity in the bistable regime. Therefore, our
results suggest that even small excitatory stimuli in a random neural network
arriving from other parts might be sufficient for the initiation of excessively
high neural synchronisation, related to the onset of epileptic seizures. Thus,
further work on other neural networks that resemble brain activity might
provide more insights on epileptogenesis.

Similarly, we apply SCP when the neurons in the network show synchronised
bursts of firing activity in the bistable regime. Here, we aim to suppressing
the synchronous behaviour by means of applying SCP. We consider SCP with
positive and negative amplitudes applied to $10\%$ of the neurons in the
random network. Figure \ref{fig5}(a) shows how long the bursts remain
synchronised after SCP is switched off (colour bar). We verified that both
negative and positive amplitudes exhibit regions where the synchronous
behaviours are suppressed, namely there is a transition from synchronised
bursts to desynchronised spikes. In addition, for $T_I>0.4$s and considering
the absolute value of the amplitudes, the transition occurs for positive values
with smaller amplitudes than for negative values. In Fig. \ref{fig5}(b), we
show the dependence of the percentage of the perturbed neurons by the stimulus
on the time the neurons remain in the bursting synchronous regime. The black
region represents parameters for which the dynamics on the network does not
remain synchronous, and therefore, synchronisation is suppressed. In this
figure, $T_I=1$s. These results allow us to conclude that desynchronous
behaviour is achieved for $A_I>15$pA and for at least $10\%$ of the perturbed
neurons.


\section{Discussion and Conclusion}

In this paper, we studied the influence of inhibitory synapses on the
appearance of synchronised and desynchronised fire patterns in a random
network with adaptive exponential integrate-and-fire neural dynamics. When the
inhibitory influence is reduced by either decreasing the inhibitory synaptic
strength or the number of inhibitory neurons, the dynamics on the network is
more likely to exhibit synchronous behaviour. The occurrence of synchronisation
results from the lack of balance between excitatory and inhibitory synaptic
influences.

We found parameter values that shift to a bistable regime where the neurons can
either exhibit desynchronous spiking or synchronised bursting behaviour. In the
bistability region, a desynchronous (synchronous) behaviour becomes synchronous
(desynchronous) by varying forward (backward) $g_{\rm exc}$. The onset of
synchronisation is thus associated with a hysteresis-loop

We showed that, in the bistable regime, synchronised bursts can be induced by
means of applying square current pulses. Our study also showed that
outside the bistable regime, square current pulses do not induce
synchronisation. Furthermore, in the bistable regime, when neurons are
synchronised, square current pulses can be used to suppress it. Positive
amplitudes of square current pulses are more effective in ceasing synchronised
bursts than negative ones. In addition, we showed that when one applies square
current pulses to less than $10\%$ of the neurons in the network, it is enough
to desynchronise the dynamics. Our work shows that a decrease of inhibition
contributes to the appearance of excessively high synchronisation, reminiscent
of the onset of epileptic seizures in the brain, thus confirming previous
experimental results and theoretical models. Both decreasing the number of
inhibitory neurons and the inhibitory strength, induce excessively high
synchronisation, related to epilepsy.

Finally, within this framework, we hypothesise that low amplitude stimuli
coming from some brain regions might be capable of inducing an epileptic
seizure manifested by high neural (abnormal) synchronisation in other brain
regions. Therefore, the work in this paper supports the common approach of the
induction of square current pulses to control or treat epileptic seizures,
since we have shown that such external perturbations not only can induce, but
more importantly can suppress synchronous behaviour in random networks with
neural dynamics.



\section*{Aknowledgments}

This study was possible by partial financial support from the following
Brazilian government agencies: Funda\c c\~ao Arauc\'aria, CNPq ($433782/2016-1$,
$310124/2017-4$, and $428388/2018-3$), CAPES, and FAPESP ($2015/50122-0$,
$2015/07311-7$, $2016/16148-5$, $2016/23398-8$, $2017/13502-5$, $2017/18977-1$,
$2018/03211-6$). We also wish to thank the Newton Fund, COFAP, and International
Visiting Fellowships Scheme of the University of Essex. We also thank IRTG for
support.



\begin{thebibliography}{100}

\bibitem{Abdullahi2017}
Abdullahi,  A. T., $\&$ Adamu, L. H. (2017). Neural network models of
epileptogenesis, {\it Neurosciences}, 22(2), 85-93.

\bibitem{Antonopoulos2016}
Antonopoulos C. G. (2016). Dynamic range in the C.elegans brain network, {\it
  Chaos}, 26(1), 1054-1500.

\bibitem{Batista2017}
Batista, C. A. S., Szezech Jr, J. D., Batista, A. M., Macau, E. E. N., $\&$
Viana, R. L. (2017). Synchronization of phase oscillators with coupling mediated
by a diffusing substance, {\it Physica A}, 470, 236-248.

\bibitem{Borges2017}
Borges, F. S., Protachevicz, P. R., Lameu, E. L., Bonetti, R. C.,  Iarosz, K.
C., Caldas, I. L., Baptista, M. S., $\&$  Batista., A. M. (2017). Synchronised
firing patterns in a random network of adaptive exponential integrate-and-fire
neuron model, {\it Neural Networks}, 90, 1-7.

\bibitem{Brette2005}
Brette, R., $\&$  Gerstner, W. (2005). Adaptive exponential integrate-and-fire
model as an effective description of neural activity, {\it Journal of
Neurophysiology}, 94, 3637-3642.

\bibitem{Chen2015}
Chen, M., Guo, D., Li, M., Ma, T., Wu, S., Ma, J., Cui, Y., Xu, P., $\&$ Yao,
Y. (2015). Critical roles of the direct GABAergic pallido-cortical pathway in
controlling absence seizures, {\it Plos Computational Biology}, 11(10),
e1004539.

\bibitem{Chen2014}
Chen, M., Guo, D., Wang, T., Jing, W., Xia, Y., Xu, P., Luo, C., Valdes-Sosa,
P. A., $\&$ Yao, D. (2014). Bidirectional control of absencd sizures by the
basal ganglia: A computational evidence, {\it Plos Computational Biology},
10(3), e1003495.

\bibitem{Cho2015}
Cho, K.-O., Lybrand, Z. R., Ito, N., Brulet, R., Tafacory, F., Zhang, L., Good,
L., Ure, K., Kernie, S. G., Birnbaum, S. G., Scharfman, H. E., Eisch, A. J.,
$\&$  Hsieh, J. (2015). Aberrant hippocampal neurogenesis contributes to
epilepsy and associated cognitive decline, {\it Nature Communications}, 6, 6606.

\bibitem{Clopath2006}
Clopath, C., Jolivet, R., Rauch, A., L\"uscher, H.-R., $\&$ Gerstner, W.
(2007). Predicting neural activity with simple models of the threshold type:
adaptive exponential integrate-and-fire model with two compartments, {\it
Neurocomputing}, 70(10-12), 1668-1673.

\bibitem{Cota2009}
Cota, V. R., Medeiros, D. C., Vilela, M. R. S. P., Doretto, M. C., $\&$ Moraes,
M. F. D. (2009). Distinct patterns of electrical stimulation of the basolateral
amygdala influence pentylenetetrazole seizure outcome, {\it Epilepsy $\&$
Behavior}, 14, 26-31.

\bibitem{DaQing2017}
DaQing, G., Chuan, X., ShengDun, W., TianJiao, Z., YangSong, Z., Yang, X., $\&$
DeZhong, Y. (2017). Stochastic fluctuations of permittivity coupling regulate
seizure dynamics in partial epilepsy, {Science China Technological Sciences},
60(7), 995-1002.

\bibitem{Danzer2017}
Danzer, S. (2017). Mossy fiber sprouting in the epileptic brain: Taking on the
Lernaean Hydra, {\it Epilepsy Currents}, 17(1), 50-51.

\bibitem{Dingledine2014}
Dingledine, R., Varvel, N. H., $\&$ Dudek, F. E. (2014). When and how do
seizures kill neurons, and is cell death relevant to epileptogenesis? {\it
Advances in Experimental Medicine and Biology}, 813, 109-122.
	
\bibitem{Engel2013}
Engel, J. Jr., Thompson, P. M., Stern, J. M., Staba, R. J., Bragin, A., $\&$
Mody, I. (2013). Connectomics and epilepsy, {\it Current Opinion in Neurology},
26, 186-194.

\bibitem{Fisher2018}
Falco-Walter, J. J., Scheffer, I. E., $\&$ Fisher, R. S. (2018). The new
definition and classification of seizures and epilepsy, {\it Epilepsy
Research}, 139, 73-79.

\bibitem{Fisher2005}
Fisher, R. S., van Emde Boas, W., Blume, W., Elger, C., Genton, P., Lee, P.,
$\&$ Engel, J. Jr. (2005). Epileptic seizures and epilepsy: {D}efinitions
proposed by the International League Against Epilepsy (ILAE) and the
International Bureau for Epilepsy (IBE), {\it  Epilepsia}, 46(4), 470-472.

\bibitem{Geier2017}
Geier, C., $\&$ Lehnertz, K. (2017). Which Brain Regions are Important
for Seizure Dynamics in Epileptic Networks? Influence of Link
Identification and EEG Recording Montage on Node Centralities,
{\it International Journal of Neural Systems}, 27, 1650033.

\bibitem{Guo2016b}
Guo, D., Chen, M., Perc, M., Wu, S., Xia, C., Zhang, Y., Xu, P., Xia, Y., $\&$
Yao, D. (2016). Firing regulation of fast-spiking interneurons by autaptic
inhibition, {\it Europhysics Letters}, 114, 30001.

\bibitem{Guo2016a}
Guo, D., Wu, S., Chen, M., Perc, M., Zhang, Y., Ma, J., Cui, Y., Xu, P., Xia,
Y., $\&$ Yao, D. (2016). Regulation of irregular neuronal firing by autaptic
transmission, {\it Scientific Reports}, 6, 26096.

\bibitem{Holt2013}
Holt, A. B., $\&$ Netoff, T. I. (2013). Computational modeling of epilepsy for
an experimental neurologist, {\it Experimental Neurology}, 244, 75-86.

\bibitem{Jessberger2015}
Jessberger, S., $\&$  Parent, J. M. (2015). Epilepsy and adult neurogenesis,
{\it Cold Spring Harbor Perspectives in Biology}, 7, 1-10.

\bibitem{Jiruska2013}
Jiruska, P., de Curtis, M., Jefferys, J. G. R., Schevon, C. A., Schiff, S.
J., $\&$ Schindler, K. (2013). Synchronization and desynchronization in
epilepsy: {C}ontroversies and hypotheses, {\it Journal Physiological}, 591.4,
787-797.

\bibitem{Khajanchi2018}
Khajanchi, S., Perc, M., $\&$ Ghosh, D. (2018). The influence of time delay in
a chaotic cancer model, {\it Chaos}, 28, 103101.

\bibitem{Kramer2012}
Kramer, M. A., $\&$ Cash, S. S. (2012). Epilepsy as a disorder of cortical
network organization, {\it Neuroscientist}, 18(4), 360-372.

\bibitem{Kuramoto1984}
Kuramoto, Y (1984). Chemical oscillations, waves, and turbulence. Berlin:
Springer-Verlag.

\bibitem{Li2007}
Li, X., Cui, D., Jiruska, P., Fox, J. E., Yao, X., $\&$ Jefferys, J. G. (2007).
Synchronization measurement of multiple neural populations, {\it J.
Neurophysiol.}, 98, 3341-3348.
  
\bibitem{Lundqvist2010}
Lundqvist, M., Compte, A., $\&$ Lansner, A. (2010). Bistable, irregular firing
and population oscillations in a modular attractor memory network, {\it Plos
Computational Biology}, 6(6), e1000803.

\bibitem{McCandless2012}
McCandless, D. W. (2012). Epilepsy: animal and human correlations. New York:
Springer-Verlag.

\bibitem{Naud2008}
Naud, R., Marcille, N., Clopath, C., $\&$ Gerstner, W. (2008). fire patterns
in the adaptive exponential integrate-and-fire model, {\it Biological
Cybernetics}, 99, 335-347.

\bibitem{noback05}
Noback, C. R., Strominger, N. L., Demarest, R. J., $\&$ Ruggiero, D. A. (2005).
\textit{The Human Nervous System: Structure and Function} (Sixth ed.). Totowa,
NJ: Humana Press.

\bibitem{Noachtar2009}
Noachtar, S., $\&$ R\'emi, J. (2009). The role of EEG in epilepsy: {A} critical
review, {\it Epilepsy $\&$ Behavior}, 15, 22-33.

\bibitem{Ostojic2014}
Ostojic, S. (2014). Two types of asynchronous activity in networks of
excitatory and inhibitory spiking neurons, {\it Nature Neuroscience}, 17,
594-600.

\bibitem{Protachevicz2018}
Protachevicz, P. R., Borges,  R. R., Reis, A. S., Borges, F. S., Iarosz, K. C.,
Caldas, I. L., Lameu, E. L., Macau,  E. E. N., Viana, R. L., Sokolov, I. M.,
Ferrari, F. A. S., Kurths, J., $\&$ Batista, A. M. (2018). Synchronous
behaviour in network model based on human cortico-cortical connections, {\it
Physiological Measurement}, 39(7), 074006.

\bibitem{Rosenblum96}
Rosenblum, M. G., Pikowsky, A. S., $\&$ Kurths, J. (1996). Phase
synchronization of chaotic oscillators, {\it Physical Review Letters}, 76(11),
1804-1807.

\bibitem{Rosenblum97}
Rosenblum, M. G., Pikowsky, A. S., $\&$ Kurths, J. (1997). From phase to lag
synchronization in coupled chaotic oscillators, {\it Physical Review Letters},
78(22), 4193-4196.

\bibitem{Scharfman2014}
Scharfman, H. E., $\&$ Buckmaster, P. S. (2014). Issues in clinical
epileptology: A view from the bench. New York: Springer.

\bibitem{Schindler2008}
Schindler, K. A., Bialonski, S., Horstmann, M. T., Elger, C. E., $\&$ Lehnertz,
K. (2008). Evolving functional network properties and synchronizability during
human epileptic seizures, {\it Chaos}, 18, 033119.

\bibitem{Sierra-Paredes2007}
Sierra-Paredes, G., $\&$  Sierra-Marcu\~no, G. (2007). Extrasynaptic GABA and
glutamate receptors in epilepsy, {\it CNS Neurological Disorders Drug Targets},
6, 288-300.

\bibitem{Silva2003}
Silva, F. H. L., Blanes, W., Kalitzin, S. N., Parra, J., Suffczynski, P., $\&$
Velis, D. N. (2003). Dynamical diseases of brain systems: {D}ifferent routes to
epileptic seizures, {\it IEEE Transactions on Biomedical Engineering}, 50,
540-548.

\bibitem{Suffczynski2004}
Suffczynski, P., Kalitzin, S., $\&$ Da Silva, F. H. L. (2004). Dynamic of
non-convulsive epileptic phenomena modeled by a bistable network, {\it
Neuroscience}, 126, 467-484.

\bibitem{Sun2018}
Sun, X., Perc, M., Kurths, J., $\&$ Lu, Q. (2018). Fast regular firings induced
by intra- and inter-time delays in two clustered neuronal networks, {\it Chaos},
28, 106310.

\bibitem{Traub1994}
Traub, R. D., Jefferyst, J. G. R., $\&$ Whittington, M. A. (1994). Enhanced
NMDA conductance can account for epileptiform activity induced by low Mg2+ in
the rat hippocampal slice, {\it Journal of Physiology}, 478, 379-393.

\bibitem{Traub1993}
Traub, R. D., Miles, R., $\&$ Jefferys, J. G. R. (1993). Synaptic and intrinsic
conductances shape picrotoxin-induced synchronized after-discharges in the
guinea-pig hippocampal slice, {\it The Journal of Physiology}, 461, 525-547.

\bibitem{Traub1982}
Traub, R. D. $\&$ Wong, R. K. S. (1982). Cellular mechanism of neural
synchronization in epilepsy, {\it Science}, 216, 745-747.

\bibitem{Trinka2015}
Trinka, E., Cock, H., Hesdorffer, D., Rossetti, A., Scheffer, I. E., Shinnar,
S., Shorvon, S., $\&$ Lowenstein, D. H. (2015). A definition and classification
of status epilepticus-report of the ILAE task force on classification of
status epilepticus, {\it Epilepsia}, 56, 1515-1523.

\bibitem{Uhlhaas2006}
Uhlhaas, P. J., $\&$ Singer, W. (2006). Neural synchrony in brain disorders:
relevance for cognitive dysfunctions and pathophysiology, {\it Neuron}, 52,
155-168.

\bibitem{Velasco2007}
Velasco, A. L., Velasco, F., Velasco, M., Trejo, D., Casto, G., $\&$
Carrillo-Ruiz., J. D. (2007). Electrical stimulation of the hippocampal
epileptic foci for seizure control: {A} double-blind, long-term follow-up
study, {\it Epilepsia}, 48(10), 1895-1903.

\bibitem{Wang2017}
Wang, Z., Tian, C., Dhamala,  M., $\&$ Liu., Z. (2017). A small change in
network topology can induce explosive synchronization and activity
propagation in the entire network, {\it Scientific Reports}, 7, 561.

\bibitem{White2002}
White, H. S. (2002). Animal models of epileptogenesis, {\it Neurology}, 59,
7-14.

\bibitem{Wong2005}
Wong, M. (2005). Modulation of dendritic spines in epilepsy: {Cellular}
mechanisms and functional implications, {\it Epilepsy $\&$ Behavior}, 7,
569-577.

\bibitem{Wong2008}
Wong, M. (2008). Stabilizing dendritic structure as a novel therapeutic
approach for epilepsy, {\it Expert Review of Neurotherapeutics}, 8(6), 907-915.

\bibitem{Wu2015}
Wu, Y., Liu, D., $\&$ Song. Z. (2015). Neural networks and energy bursts in
epilepsy, {\it Neuroscience}, 287, 175-186.

\end{thebibliography}
\end{document}